\documentclass[aps,floatfix,prb,superscriptaddress,twocolumn]{revtex4}
\usepackage{bbm}
\usepackage{graphicx}
\usepackage{dcolumn}
\usepackage{bm}
\usepackage{subfigure}
\usepackage{amsmath}
\usepackage{amsfonts}
\usepackage{feynmf}
\usepackage{hyperref}
\newcommand{\ve}[1]{\mbox{\boldmath$#1$}}

\usepackage{attachfile} 
\newcommand{\ua}{\uparrow} 
\newcommand{\da}{\downarrow} 
\newcommand{\ra}{\rightarrow}

\newcommand{\bs}{\boldsymbol}

\newcommand{\bk}{\boldsymbol k}

\usepackage{times}

\begin{document}
\title{Edge current and orbital angular momentum of chiral superfluids revisited}
\author{Wenxing Nie}
\affiliation{Center for Theoretical Physics, College of Physics, Sichuan University, Chengdu 610064, China}
\author{Wen Huang}
\email{huangw3@sustech.edu.cn} 
\affiliation{Shenzhen Institute for Quantum Science and Engineering, Guangdong Provincial Key Laboratory of Quantum Science and Engineering, Southern University of Science and Technology, Shenzhen 518055, Guangdong, China}
\affiliation{Institute for Advanced Study, Tsinghua University, Beijing 100084, China}
\author{Hong Yao}
\email{yaohong@tsinghua.edu.cn}
\affiliation{Institute for Advanced Study, Tsinghua University, Beijing 100084, China}
\affiliation{ State Key Laboratory of Low Dimensional Quantum Physics, Tsinghua University, Beijing 100084, China }
\date{\today}

\begin{abstract}
Cooper pairs in chiral superfluids carry quantized units of relative orbital angular momentum (OAM). Various predictions of the intrinsic OAM density or the macroscopic OAM of a two-dimensional chiral superfluid differ by several orders of magnitude, which constitute the so-called {\it Angular Momentum Paradox}. Following several previous studies, we substantiate the semiclassical Bogoliubov-de Gennes theory of the {\it single-particle} edge current and OAM in two-dimensional chiral superfluids in the BCS limit. The analysis provides a simple intuitive understanding for the vanishing of OAM for a non-p-wave chiral superfluid (such as $d+id$) confined in a rigid potential. When generalized to anisotropic chiral superconductors and three-dimensional chiral superfluids, the theory similarly returns an accurate description. We also present a detailed numerical study of the chiral phases in the BEC limit. Our study suggests that, in both BCS and BEC phases the relative OAM of the individual Cooper pairs contribute to the total OAM additively, and that in both phases the corresponding macroscopic OAM density distribution is localized at the boundary. 
\end{abstract}
\maketitle
\section{Introduction}
\label{sec:Introduction}
A chiral superfluid is one in which each Cooper pair carries a quantized relative orbital angular moment (OAM)~\cite{Anderson:61,Leggett:75}, e.g. $L_z= \nu\hbar$ where $\nu=1,2,3$ for chiral p-, d- and f-waves in two dimensions (2D). In the Nambu spinor basis $\psi_{\bk}=(c_{\bk,\ua},c^\dagger_{-\bk,\da})^T$, the Bogoliubov-de Gennes (BdG) Hamiltonian reads,
\begin{equation}
H_{\bk} = \psi_{\bk}^\dagger \begin{bmatrix}
\xi_{\bk}  & \Delta  e^{i \nu\theta_{\bk}} \\
\Delta e^{-i\nu\theta_{\bk}} & -\xi_{-\bk} \end{bmatrix}  \psi_{\bk},
\label{eq:Hamiltonian}
\end{equation}
where $\xi_{\bk} = k^2/2m-E_f$ is the normal state band dispersion, $\theta_{\bk}$ denotes the direction of the wavevector $\bk$, and the gap function $\Delta_{\bk}=\Delta e^{i \nu\theta_{\bk}} = \Delta(k_x+ik_y)^\nu/k^\nu$ encrypts the chirality of the pairing. Note that the pairing is a spin-triplet for $\nu$ odd and a spin-singlet for $\nu$ even. 

One natural question is whether a chiral superfluid exhibits an overall OAM, and if yes, how such a macroscopic quantity relates to the relative OAM carried by the individual Cooper pairs. The question was originally raised for the Anderson-Brinkman-Morel phase (A-phase) of $^3$He~\cite{Anderson:61,Anderson:73,Leggett:75} -- which is a chiral p-wave superfluid in the BCS limit with $|\Delta| \ll E_f$, and it has been a subject of long-standing controversy~\cite{Leggett:75,LeggettBook,Mizushima:16}. We refer to Ref.~\onlinecite{Mizushima:16} for a recent thorough review of the theoretical developments over the past four decades. In short, predictions of the intrinsic OAM density in the bulk of the superfluid vary from $L_z^\text{tot} = (\Delta/E_f)^2 \rho \hbar/2$ to $\rho \hbar/2$, where $\rho$ is the particle density; and likewise for the total OAM carried by a finite $N$-particle system by replacing $\rho$ with $N$~\cite{Anderson:61,Leggett:75,Volovik:75,Cross:77,Mermin:75,Ishikawa:77,McClure:79,Mermin:79,Volovik:81,Kita:96}. A logical deduction could be made for non-p-wave chiral superfluids simply by multiplying the above quantities by a corresponding $\nu$. Such predictions span over many orders of magnitude, constituting the celebrated {\it Angular Momentum Paradox}~\cite{Leggett:75,LeggettBook,Mizushima:16}.

\begin{figure}
\includegraphics[width=8.5cm]{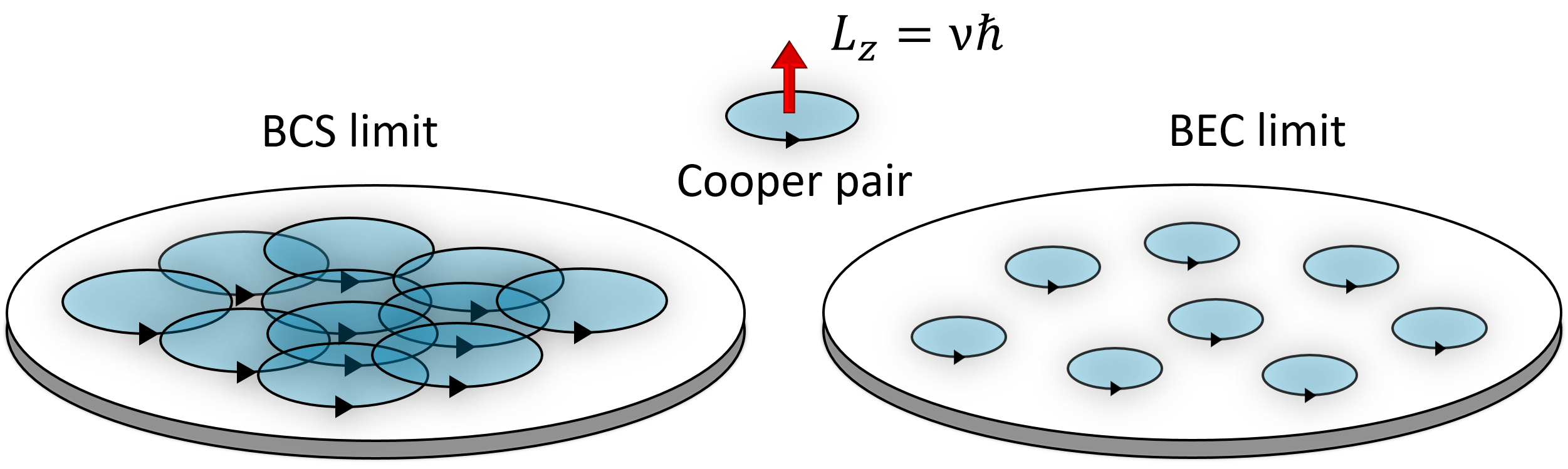}
\caption{Illustration of 2D BCS and BEC chiral superfluids. The average Cooper pair size in a BCS superfluids is much larger than the interparticle spacing, hence such superfluids are typically pictorially depicted as strongly overlapping Cooper pairs. By contrast, the BEC limit is routinely portrayed as tightly-bound Cooper pair molecules spread evenly across the system. A central quest of the {\it angular momentum paradox} is the total OAM carried by the chiral superfluids confined in a disk. }
\label{fig:illustration}
\end{figure}

A separate, and important, aspect of the theory of the BCS chiral superfluids is their nontrivial topological property, which is classified by an integer topological invariant \cite{Volovik:88} -- the Chern number $C$, that coincides with the Cooper pair angular momentum, i.e. $C=\nu$. The same number of chiral edge dispersion, and therefore spontaneous chiral edge current, may emerge at the boundary of the system. This has been a subject of a certain amount of confusion. In particular, it had been tempting to (erroneously) relate the edge current and OAM to the underlying topology. 

Multiple pioneering studies on the front of chiral p-wave have appeared since the turn of the century \cite{Matsumoto:99,Furusaki:01,Stone:04,Stone:08,Sauls:11}. In particular, on the basis of mean-field BdG calculations, the total OAM of a system in a confining geometry was found to coincide with $N\hbar/2$ \cite{Stone:04,Stone:08,Sauls:11} -- the intuitively expected value obtained by assuming that all particles are involved in Cooper pairing and that each pair contributes $L_z = \hbar$. Within this formalism, the OAM with respect to the center of the geometry is generated entirely by the spontaneous edge current. In this regard, the OAM thus obtained is a single-particle quantity: the spontaneous current originates from single-particle surface scattering, which takes place against the backdrop of a Cooper pairing with a specific chirality. Spatially, the bulk of the geometry has no contribution at all. Therefore, this OAM must be distinguished from the OAM density defined as a two-particle pair correlation~\cite{Anderson:61,Leggett:75}, which is intrinsically sensitive to the pair condensation amplitude and which distributes uniformly across the bulk of the system. We will not elaborate their distinction in the present study. 

Later, two simultaneous and somewhat differently formulated studies within the BdG framework showed unambiguously that the total edge current and OAM must effectively vanish for any 2D non-p-wave BCS chiral superfluids in sharp confining potentials~\cite{Huang:14,Tada:15}, despite their nontrivial topological ground states, and despite the higher relative OAM each Cooper pair carries. This seemingly counter-intuitive result is supported by several subsequent studies~\cite{Volovik:14,Ojanen:16,Suzuki:16,WangX:18}. The remarkable conclusion also applies to $s$-wave superfluids carrying multiple vortices~\cite{Prem:17}. Within the formulation of Ref.~\onlinecite{Tada:15}, the total OAM has two distinct origins, one is the relative OAM between paired fermions (i.e. $\nu\hbar$ from each pair), and the other is the OAM carried by unpaired fermions at the boundary. Crucial information is encapsulated in the many-body ground state wavefunction in the sharp confinement. For a chiral p-wave superfluid, all particles are involved in Cooper pairing and thus a total OAM of $N\hbar/2$ is obtained; for non-p-wave states, however, some fermions are unpaired and they carry a net OAM which (in the limit $\Delta/E_f\to0$) essentially cancels the contribution from the remaining paired particles. Notably, the presence of unpaired fermions at the boundary of a finite-size geometry, although only relevant for non-p-wave states, was not recognized in most previous literature~\cite{Ishikawa:77,McClure:79,Kita:96}. On the other hand, although the semiclassical analyses in Ref.~\onlinecite{Huang:14} is intuitive and solid, some important details were not made sufficiently explicit. In this work, we substantiate those analyses for the case of rigid confinement and corroborate with numerical BdG calculations when necessary. Generalizing these analyses to 3D chiral superfluids also yields accurate descriptions. These conclusions are all in qualitative agreement with a phenomenological Ginzburg-Landau theory, within which the current is predominantly associated with terms describing the correlations of the spatial variation of the different order parameter components in orthogonal spatial directions \cite{Volovik:85,Sigrist:89,Furusaki:01,Bouhon:14,Huang:14}. 

The relation to topology was also further elucidated in Refs.~\onlinecite{Huang:15,Tada:15b}. In essence, due to $U(1)$ symmetry breaking, the spontaneous current and the OAM are not topologically protected quantities and thus are not directly related to the Chern number, a fact which should have already been anticipated by accounting for the lack of a genuine Chern-Simons action \cite{Volovik:88,Furusaki:01,Stone:04,Huang:15}. This paves the way for explaining the absence or smallness of the edge current \cite{Kirtley:07,Hicks:10,Curran:14} in the putative time-reversal symmetry breaking superconductor Sr$_2$RuO$_4$ by invoking gap anisotropy and/or surface disorder \cite{Ashby:09,Lederer:14,Bouhon:14,Huang:15,Tada:15b,Scaffidi:15,Tada:18}.

Finally, in contrast to the BCS limit where the average Cooper pair size is much larger than the interparticle spacing, the BEC limit with $E_f<0$ is routinely portrayed as a macroscopic coherent state of tightly-bound Cooper pair molecules (see Fig.~\ref{fig:illustration}). Besides this, BEC chiral superfluids are topologically trivial with $C=0$. As such, one prevailing understanding has been that their total OAM should be $\nu N\hbar/2$ and that it should be uniformly distributed across the system. Numerical BdG calculations indeed obtained the expected OAM~\cite{Tada:15}. In this study, we investigate the real space distribution of the edge current and the associated OAM. We find again that they arise only at the boundary, as oppose to the expectation stated above. 

The rest of the paper is organized as follows. In Sec.~\ref{sec:2Dchiral}, we revisit the semiclassical analysis of the edge current and the OAM in the BCS limit and in the presence of rigid confining potentials, supplementing rigorous derivations for some important details. The same analysis is extended to 2D chiral superconductors on a lattice, and then in Sec.~\ref{sec:3Dchiral} to 3D chiral superfluids, along with extensive numerical BdG calculations in support of the conclusions. In Secs.~\ref{subSec:Alternative} and~\ref{subsec:SoftEdge}, we comment on two parallel theories, i.e. the Gingzburg-Landau theories of the spontaneous current and the spectral flow argument for the OAM, and briefly discuss the scenario with soft confining potential and comment on the significance of the non-topological Chern-Simons-like action. Going to the BEC limit in Sec.~\ref{sec:2Dbec}, numerical calculations show that the OAM is also confined to the boundary as in the BCS limit, which runs contrary to common beliefs. The paper is briefly summarized in Sec.~\ref{sec:summary}.

\section{2D chiral superfluids in BCS limit}
\label{sec:2Dchiral}
In this section, we revisit a semiclassical analysis which intuitively explains the vanishing of edge current and OAM in non-p-wave chiral superfluids and the suppression thereof in anisotropic chiral p-wave superconductors. Spirits of what follows have appeared in a number of previous literature \cite{Furusaki:01,Stone:04,Huang:14,Huang:15,Suzuki:16}, in particular Refs.~\onlinecite{Huang:14,Suzuki:16} for non-p-wave chiral states. We shall substantiate the relevant arguments with more rigorous derivations and state them in more explicit languages. Later we shall comment on parallel approaches, i.e. the Gingzburg-Landau theory and the spectral flow argument \cite{Volovik:95,Volovik:14,Tada:15}. 

\subsection{Semiclassical theory in continuum limit}
In the following derivation, it turns out beneficial to remove the $k$-dependence in the denominator of the gap function in Eq.~(\ref{eq:Hamiltonian}). We hence take $\Delta_{\bk} = \Delta(k_x+ik_y)^\nu/k_f^\nu$, where $k_f$ represents the Fermi wavevector. Although this changes the gap amplitude,  the global topological nature of the pairing, henceforth the essential properties of the chiral edge modes, are retained so long as the relation $\Delta/E_f \ll 1$ is satisfied, as has been demonstrated in previous literature~\cite{Stone:04,Sauls:11}. Hence the choice of $k_f$ is not special but contingent. 

\begin{figure}
\includegraphics[width=8.5cm]{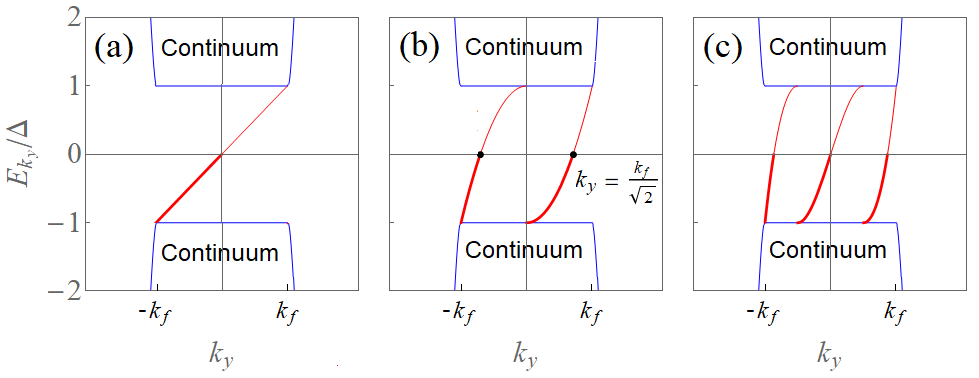}
\caption{Sketch of chiral edge modes (red) in the BCS limit (a) chiral p-wave, (b) chiral d-wave, and (c) chiral f-wave states in a half-infinite plane with a boundary parallel to the $y$-axis. The thickened segments correspond to negative-energy edge modes that are occupied in the ground state. In chiral d- and f-wave models, there are more than one chiral edge branch, however, the current carried by the multiple branches of {\it occupied} edge modes cancel each other in the limit $\Delta/E_f \ra 0$.}
\label{fig:edgeDisp}
\end{figure}

In a half-infinite geometry with an ideal sharp boundary parallel to the $y$-axis, the edge dispersion, e.g. for chiral p- and d-waves, acquires the following form (see App. \ref{app1} and Fig. \ref{fig:edgeDisp}),
\begin{eqnarray}
\text{p-wave}:~E_{k_y} &=& \Delta k_y/k_f\,, \nonumber \\
\text{d-wave}:~E_{k_y} &=& \left\{ \begin{array}{cc} 
                \Delta(k_f^2-2k_y^2)/k_f^2\,, & ~k_y\in(-k_f,0) \\
                -\Delta (k_f^2-2k_y^2)/k_f^2\,, & ~k_y\in(0,k_f) \\
                \end{array} \right.  \nonumber \\
\label{eq:PDdisp}
\end{eqnarray}
with the wavefunction given by,
\begin{equation}
\phi_{k_y}(x,y)= \frac{1}{\mathcal{N}}\begin{pmatrix}
u_{k_y} \\
v_{k_y}
\end{pmatrix} \sin(k_{fx}x)e^{-\frac{\Delta}{v_f} x}e^{ik_{fy}y}  \,.
\end{equation}
Here $\mathcal{N}$ is a normalization factor, $\bs k_f = (k_{fx},k_{fy})$ the Fermi wavevector, $v_f=k_f/m$, and importantly, $(u_{k_y},v_{k_y})^T = (1,-i)^T/\sqrt{2}$ for p-wave and $(u_{k_y},v_{k_y})^T=(1,\pm 1)^T/\sqrt{2}$ for d-wave where the plus and minus signs are associated respectively with its two chiral branches. As we elaborate in App.~\ref{app1}, the equal-weight particle and hole composition of the edge modes in chiral p-wave is a consequence of a chiral symmetry. The non-p-wave superfluids, however, lacks such a symmetry, and $u_{k_y}$'s and $v_{k_y}$'s typically exhibit a correction of order $\mathcal{O}(\Delta/E_f)$ (App.~\ref{app1}). In the BCS limit with $\Delta/E_f \ll 0$, these edge states, described by operators $\gamma_{k_y} =( u_{k_y} c_{k_y,s} + v_{k_y} c^\dagger_{-k_y,s^\prime})$, are nonetheless essentially `{\it charge-neutral}' (importantly, they nevertheless carry a finite amount of current. See below). 

We note that the matrix in Eq.~(\ref{eq:Hamiltonian}) has the same mathematical structure as that of a Chern insulator (CI), which is characterized by the same topological invariant and therefore the same edge dispersion. However, the two topological states differ in important ways. Unlike a CI which preserves the charge $U(1)$ symmetry, the low-energy effective action of the chiral superfluid lacks a real Chern-Simons term that embodies a protected quantized particle current at the boundary of the system \cite{Stone:04}. Hence the edge current of a topological chiral superfluid is bona fide non-topological \cite{Huang:15,Tada:15}. Formally, this distinction manifests in their velocity operators, i.e. $\hat{\mathcal{V}}_k = \partial_k \xi_k \sigma_0$ for chiral superfluids and $\hat{\mathcal{V}}_k = \partial_k H_k$ for CIs, which foretells profound consequences. In particular, the current carried by an individual bogoliubov quasiparticle is {\it entirely} unrelated to its group velocity!

Focusing on the chiral edge states, at zero temperature the particle current carried by an individual occupied edge state with characteristic momentum $k_y$ is given by,
\begin{equation}
j_{k_y} = |u_{k_y}|^2 \partial_{k_y} \xi_k = \frac{k_y}{2m} \,.
\label{eq:Jky}
\end{equation}
Lattice generalization of this expression is separately verified in our numerical BdG calculations on both chiral p- and d-wave models, up to a $\mathcal{O}(\Delta/E_f)$ correction when it is present (see App.~\ref{app:nuBdG}). We checked that the correction (when present) has nothing to do with the group velocity of the edge mode, but is purely a consequence of the above stated correction to the quasiparticle wavefunction $(u_{k_y},v_{k_y})^T$.

The particle current generates a mass current, i.e. linear momentum, of $k_y/2$. It is straightforward to evaluate the total edge state contribution (per spin species, same hereafter), $J_\text{e}= \frac{1}{2\pi}\int^\prime j_{k_y} dk_y $, where the prime indicates integration over the occupied edge modes. Inspection of the edge dispersion in Fig.~\ref{fig:edgeDisp} (a) reveals that, $J_\text{e}$ is always positive (or negative) in the p-wave case. However, the peculiar momentum-space distribution of the edge dispersion in chiral d-wave [Fig.~\ref{fig:edgeDisp} (b)] implies that contributions from the two chiral branches flow in opposite directions. It can in fact be further shown that the spatially integrated current carried by the multiple chiral edge branches perfectly cancel~\cite{Huang:14}, not only for d-wave, but also for f-wave and all other non-p-wave chiral superfluids in two spatial dimensions. The continuum states may also carry a net particle current $J_\text{e}$ localized to the boundary, and the total edge current follows as $J_\text{tot}=J_\text{e}+J_\text{c}$. Spatially, the currents $J_\text{e}$, $J_\text{c}$, and thus $J_\text{tot}$, are all localized at the boundary over superfluid coherence length scales. However, except in the p-wave case where the continuum contribution is half of $J_\text{e}$ and in opposite direction~\cite{Stone:04} [which also holds for simple lattice models below, such as shown in Fig.~\ref{fig:PwaveSpecFig} (a) and (d)], $J_\text{c}$ vanishes identically for all non-p-wave states~\cite{Huang:14}. In other words, these states have $J_\text{tot}=J_\text{e}=J_\text{c}=0$. This signifies, rather strikingly, vanishing total OAM for any non-p-wave chiral superfluid placed on a disk much larger than the coherence length \cite{Huang:14,Tada:15}! While for chiral p-wave~\cite{Stone:04,Sauls:11}, 
\begin{equation}
J_\text{e}=\frac{k_f^2}{8\pi m} =\frac{\rho}{2m}\,, \text{and}~~~ J_\text{c}=-\frac{\rho}{4m} \,,
\end{equation}
where $\rho$ represents the particle density (per unit area), which remains constant throughout the system except for some inessential short-wavelength Friedel oscillations near the boundary. The total edge current is then given by,
\begin{equation}
J_\text{tot} = \frac{\rho}{4m}\,.
\end{equation}
In a disk geometry with large radius $R$, such edge current generates a net OAM~\cite{Stone:04,Sauls:11}, 
\begin{equation}
L^\text{tot}_z = mJ_\text{tot}R\cdot 2\pi R = \frac{N\hbar}{2}\,,
\end{equation}
where $N=\pi R^2\rho$ is the total number of particles in the disk. This result coincides with the intuitive expectation for chiral p-wave superfluids. 

Noteworthily, although the spatially integrated current vanishes for non-p-wave states, counter flowing local currents distribute within a region comparable to a coherence length from the edge~\cite{Braunecker:05,Suzuki:16,WangX:18}. Furthermore, in typical numerical calculations, additional superconducting order parameters may emerge due to symmetry breaking at the edge, resulting in small nonvanishing integrated current~\cite{WangX:18}. As a final note, in a CI the current of an individual mode is set by $\partial_{k_y} E_{k_y}$, and is therefore intimately tied to the chirality, i.e. the Chern number, hence the topological protection.  

\subsection{Lattice models}
\begin{figure}
\includegraphics[width=8.5cm]{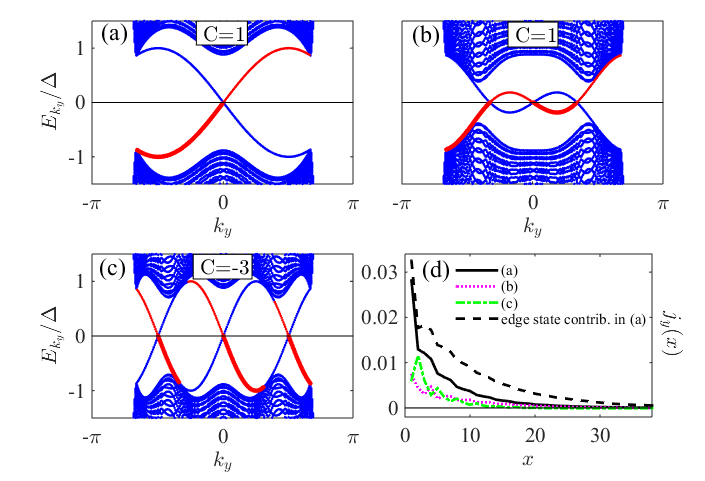}
\caption{(a)-(c) Low-energy dispersion of three 2D chiral p-wave models on square lattice in a cylindrical geometry with open boundaries in the $x$-direction. The model contains only nearest neighbor hopping $t$: $\xi_k = -2t(\cos k_x + \cos k_y)-\mu$ with $\mu=-1$. The gap function takes the following forms: (a) $\Delta_k=\Delta(\sin k_x + i \sin k_y)$; (b) $\Delta_k=\Delta(\sin k_x \cos k_y + i \cos k_x\sin k_y)$; and (c) $\Delta_k=\Delta(\sin 2k_x + i \sin 2k_y)$; and $\Delta=0.1t$. The Chern number in each case is displayed in boxes. The chiral edge modes on one of the edges are marked in red, and the thickened segments denote the occupied edge states in the ground state. (d) The edge current distribution $j_y$ for the three cases in (a)-(c). Note that the currents are expressed in units of $t$ (same in other figures). The black dashed curve shows the current carried by the occupied edge states in (a). The net edge state contribution is twice as much as the total current. The sign of $j_y$ is inverted for (b) and (c) to make a comparison with (a) more transparent. It is worth noting that for anisotropic p-waves, some portions of the edge dispersion almost merge with the bulk continuum.}
\label{fig:PwaveSpecFig}
\end{figure}

The same line of argument also explains why anisotropic chiral p-wave superconductors on lattices, when exhibiting multiple accidental zero-crossings in the edge spectrum, may support a much reduced current compared to that of a simple isotropic p-wave superconductor. This is demonstrated in Fig.~\ref{fig:PwaveSpecFig} for calculations on a square lattice with only nearest neighbor hopping, using three different p-wave gap functions that increase in levels of gap anisotropy from Fig.~\ref{fig:PwaveSpecFig} (a) to (c). The details of the calculation are presented in App.~\ref{app:nuBdG}. At the indicated chemical potential, only one chiral dispersion appears for $\Delta_{\bk}  \propto \sin k_x + i \sin k_y$ and $\sin k_x \cos k_y + i \cos k_x \sin k_y$, where the Chern number $C=1$. For an edge parallel to $y$-axis, the chiral edge dispersion is related to the $y$-component of the gap function by $|E_{\bk}| = |\Delta_{\bk y}|$ (see App.~\ref{app1}). Notably, the latter state possesses additional zero-crossings away from $k_y=0$. The number of chiral branches increases to 3 for $\Delta_k \sim \sin 2k_x + i \sin 2k_y$, where $C=-3$. Since each edge mode carries a current $j_{k_y}=\partial_{k_y}\xi_k/2 = t\sin k_y$, the multiple zero crossings in anisotropic pairings lead to partial cancellation between the current carried by different occupied edge states. As a result, the total edge current, including the continuum-state contributions, is in general reduced for anisotropic pairings \cite{Bouhon:14,Huang:15,Scaffidi:15} [see Fig.~\ref{fig:PwaveSpecFig} (d)]. 

We stress that the reduction of the spontaneous current is not solely determined by the Chern number, as it can happen even for the case of $C=1$ in Fig.\ref{fig:PwaveSpecFig} (b). In fact, the sign of the Chern number is equally unimportant. Simply inverting the normal state band dispersion $\xi_k \ra -\xi_k$ in the continuum model or setting $\mu \ra -\mu$ in the lattice model above (while keeping $\Delta_k$ unchanged), changes the sign of $C$ and inverts the chirality of the edge dispersion. However, exactly the same edge current (sign and magnitude) follows from our analyses~\cite{Huang:15}. These underscore again the non-topological nature of the spontaneous current. Nonetheless, it generally takes a fine-tuned pairing function to make the total current vanish~\cite{Huang:15}.  

\subsection{Alternative theories}
\label{subSec:Alternative}
As is well understood within the phenomenological Ginzburg-Landau theory~\cite{Volovik:85,Sigrist:89,Furusaki:01,Huang:14,Zhang:18,Etter:18}, the spontaneous current is closely related to the distinct textures acquired by the different chiral order parameter components near the boundary. Take chiral p-wave superfluid as an example, by symmetry the two components in the gap function $\Delta_k =\Delta_{1k} + i\Delta_{2k} =\Delta_1 k_x + i \Delta_2 k_y$ shall develop different textures at a generic boundary~\cite{footnote1}. For instance, since a reflection perpendicular to the $x$-direction takes $\Delta_1$ to $-\Delta_1$, this component must drop to zero over certain healing length near an edge parallel to $y$. The leading order contribution to the local current is given by $j_y(x) \sim
K_{xy}[(\partial_x \Delta_1)\Delta_2 - \Delta_1(\partial_x \Delta_2)]$. The phenomenological coefficient $K_{xy}$ sets the scale of this contribution. Following a standard free energy gradient expansion, $K_{xy} \propto \langle \partial_{k_x}\xi_k  \partial_{k_y}\xi_{k}\Delta_{1k} \Delta_{2k} \rangle_{FS}$ where $\langle ...\rangle_{FS}$ designates an average over the Fermi surface. The coefficient therefore is determined by both the microscopic details of the gap function and the underlying band structure~\cite{Bouhon:14,Huang:14,Huang:15}. In particular, it can be checked that $K_{xy}=0$ for any non-p-wave 2D chiral superfluids~\cite{Huang:14}. The same analysis applies to the 3D models in Sec.~\ref{sec:3Dchiral}. 

A parallel spectral flow argument focuses on the generalized angular momentum~\cite{Volovik:95,Tada:15} defined as $ \hat{Q} \equiv \hat{L}^\text{tot}_z - \frac{\nu}{2} \hat{N}\hbar$. Despite $\hat{L}^\text{tot}_z$ and $\hat{N}$ individually not commuting with the BdG Hamiltonian in Eq.~\eqref{eq:Hamiltonian} due to $U(1)$ symmetry breaking, $\hat{Q}$ does. A detailed analysis finds the ground state expectation value of $\hat{Q}$, $\langle Q \rangle$, to be zero for any BEC chiral superfluid, suggesting that $L^\text{tot}_z= \nu N \hbar/2$. In the BCS regime, however, $\langle Q \rangle=0$ holds only for chiral p-wave, and $\langle Q \rangle \simeq -  \nu N/2$ for any non-p-wave states up to a $\mathcal{O}(\Delta/E_f)$ correction. In other words, while a chiral p-wave superfluid has $L^\text{tot}_z=N\hbar/2$, any non-p-wave BCS chiral superfluid has an OAM which greatly reduces to $N\times \mathcal O(\Delta/E_f)$. Such a fundamental difference originates from the spectral flow, brought about by the presence of multiple zero-crossings in the edge dispersion -- which coincides with the presence of unpaired fermions in the many-body BCS ground state wavefunction~\cite{Tada:15}. Interestingly, within this description, the relative OAM each individual Cooper pair carries, $\nu\hbar$, coincides with the very OAM this pair generates with respect to the center of the disk geometry.

\subsection{soft boundary conditions}
\label{subsec:SoftEdge}
Following Refs.~\onlinecite{Huang:14,Huang:15}, we shall remark on the case of soft boundary potentials. Although a genuine Chern-Simons action is absent, a non-topological Chern-Simons-like term emerges in the phenomenological hydrodynamic action \cite{Mermin:79,Volovik:88,Stone:04,Furusaki:01,Goryo:99,Roy:08,Lutchyn:08,Hoyos:14}, $\mathcal{L}^\text{CS-like} = \tilde{C}/(8\pi) \epsilon_{0ij}A_0 \partial_i A_j$, where $\tilde{C} = C[1 + \mathcal{O}(\Delta/E_f)^2]$. Note that, in comparison to the standard Chern-Simons action, this action lacks the term $\epsilon_{i0j}A_i\partial_0 A_j$. It nevertheless generates a Hall-like spontaneous current perpendicular to the direction of spatial scalar potential variation, 
\begin{equation}
\bs j (\bs r)\simeq -\frac{C}{8\pi}  \hat{\bs z} \times \bs\nabla A_0(\bs r) = \frac{C}{4}  \hat{\bs z} \times \bs\nabla \rho(\bs r) \,,
\label{eq:CScurrent}
\end{equation}
where $A_0(\bs r)$ is the confining potential. In the second equation, the gradient of the scalar potential has been translated to that of the particle density, which is valid in 2D models. The action therefore effectively describes a current-density correlation, unlike the Chern-Simons action which is associated with a current-current correlation. A semi-quantized OAM $L_z^\text{tot} = \nu N \hbar/2$ is then recovered for a chiral superfluid confined by a circularly symmetric soft potential \cite{Huang:14,Huang:15}. Equation~(\ref{eq:CScurrent}) can be intuitively interpreted as follows. While the circulating current carried by the individual and strongly overlapping Cooper pairs cancel each other in a uniform macroscopic BCS coherent state, a residual overall current arises in the presence of a spatial density inhomogeneity. It then naturally follows that such a current depends on both the Cooper pair OAM and the particle density gradient, as in Eq.~(\ref{eq:CScurrent}). An alternative explanation rests upon the spectral flow induced as the edge potential is softened from the rigid confinement limit. We refer to Refs.~\onlinecite{Ojanen:16,Tada:18} for more details.  

Some caution is needed here. The Chern-Simons-like action is applicable only in the strict long-wavelength limit where the confining potential, equivalently the particle density, varies slowly on a length scale much longer than the Cooper pair coherence length. Notably, this term represents a contribution distinct from the one generated by the order parameter textures~\cite{Furusaki:01,Huang:15} mentioned above. These two contributions shall in general coexist near the boundary of a soft confinement. As such, $L_z^\text{tot}$ may still deviate from the semi-quantized value~\cite{Ojanen:16}, except for the chiral p-wave superfluid where a lack of spectral flow forbids any change in $L_z^\text{tot}$ as the edge potential is deformed~\cite{Volovik:14,Huang:15,Tada:15} (although local current distribution will still modify during the process).

\section{3D chiral superfluids in the BCS limit}
\label{sec:3Dchiral}
\begin{figure}
\subfigure{\includegraphics[width=3.7cm]{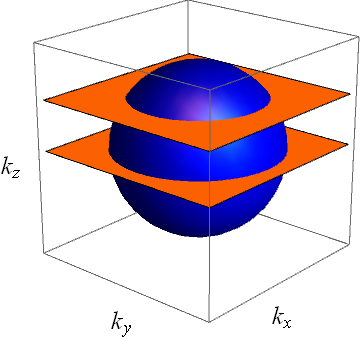}}
\subfigure{\includegraphics[width=4.8cm]{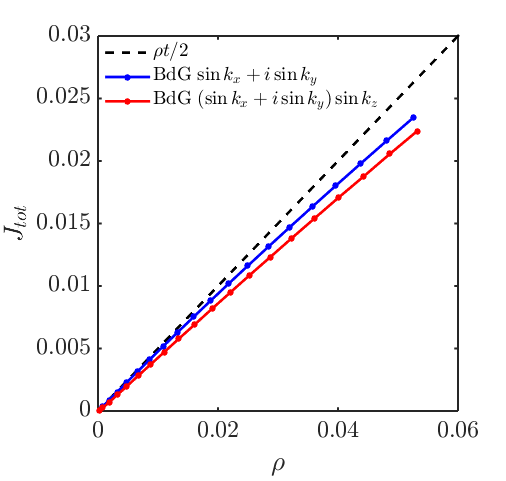}}
\caption{Left: sketch of the 3D Fermi surface and constant $k_z$ planes intersecting the Fermi surface. Right: total edge current as a function of carrier density $\rho$ in the low-density (continuum) limit, obtained from numerical BdG calculations of a cubic lattice with open boundary condition in one direction and periodic boundary conditions in other directions. We assume only nearest neighbor hopping $t$ on the lattice, with chiral p-wave $\Delta_{\bs k}= \Delta (\sin k_x + i \sin k_y)$ where $\Delta=0.1t$, and with chiral d-wave pairing $\Delta_{\bs k}= \Delta (\sin k_x + i \sin k_y)\sin k_z$ where $\Delta=0.2t$. The density $\rho$ is measured in the bulk, as the edge exhibits Fridel oscillations. The black dashed curve shows the expectation based on Eq.~(\ref{eq:semiCurr}), where we have substituted the particle mass by its lattice approximation $m\simeq 1/(2t)$ at low filling. Note that calculations of the present chiral $d$-wave model always suffer from finite size effect due to the presence of nodal quasiparticle excitations. Aside from this, its net edge current approaches that of the chiral p-wave model. }
\label{fig:3DFS}
\end{figure}

Following the above semiclassical analyses, we now turn to 3D chiral superfluids. For convenience, we dissect the 3D Brillouin zone into individual $k_z$-planes, as sketched in Fig.~\ref{fig:3DFS} (a). The Hamiltonian at each $k_z$ plane now constitutes a problem of 2D chiral pairing with an effective $k_z$-dependent Fermi energy $E_f^\prime=E_f-k_z^2/2m$. To set the stage, we consider a sharp $yz$ surface. The spontaneous current, if any, shall flow along $y$. Following the above analyses, the edge states carry a total current, 
\begin{equation}
J_\text{e}= \int_{-k_f}^{k_f}  \int_0^{\sqrt{k_f^2-k_z^2}} \frac{k_y}{2m} \frac{dk_y}{2\pi}  \frac{dk_z}{2\pi} = \frac{k_f^3}{12m\pi^2} \,.
\end{equation}
Note the $k_z$-integration is restricted to $|k_z|\leq k_f$, because beyond this regime no Fermi surface cross-section exists and hence contribution from those $k_z$-planes is negligible in comparison. Employing the conclusion that the bulk state contribution satisfies $J_\text{c}= -J_\text{e}/2$, we obtain the total current,
\begin{equation}
J_\text{tot}= \frac{k^3_f}{24m\pi^2} = \frac{\rho}{4m}
\label{eq:semiCurr}
\end{equation}
where $\rho = k_f^3/(6\pi^2)$ is the particle density (per unit volume). This is consistent with our lattice BdG calculations when approaching the continuum limit, as we demonstrate in Fig.~\ref{fig:3DFS} (b). Consider now confining this superfluid in a cylindrical container with radius $R$ and height $L$ much greater than the superfluid coherence length, the total OAM carried by the edge current is then,
\begin{equation}
L_z^\text{tot} = \frac{1}{2} \rho \pi R^2 L  = \frac{N}{2} \,,
\end{equation}
where $N=\rho \pi R^2 L$ is the total number of particles in the cylinder. This is the same as that obtained for a 2D chiral p-wave. We note that, by contrast, the thermal Hall conductivity does not follow such a simple generalization going from 2D to 3D~\cite{Yoshiaka:18}.

\begin{figure}
\subfigure{\includegraphics[width=3.7cm]{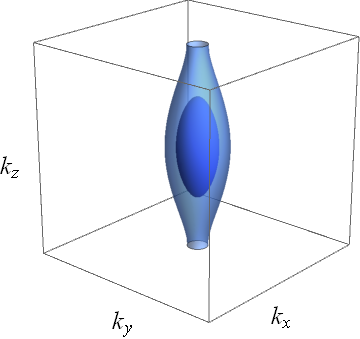}}
\subfigure{\includegraphics[width=4.8cm]{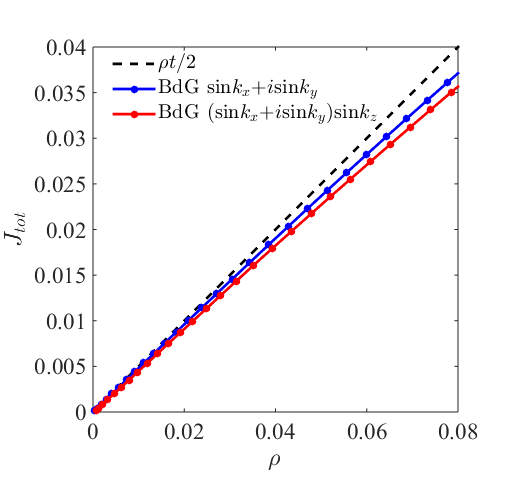}}
\caption{Left: Fermi surface contours of the model at two different filling fractions. Right: total edge current as a function of carrier density $\rho$ for the same chiral p- and d-wave pairings as in Fig. \ref{fig:3DFS}, on a 3D tetragonal lattice with out-of-plane n.n. hopping $t_z$ differing from the in-plane ones, $t_{z}=0.2t$. As in Fig.~\ref{fig:3DFS}, the black dashed line shows the expectation derived from Eq.~(\ref{eq:semiCurr}), where the particle mass is replaced by its lattice approximation $m\simeq 1/(2t)$.}
\label{fig:3DFSani}
\end{figure}

Finally, so far as the current and the OAM are concerned, the result depends only on the projection of the Cooper pair relative OAM $L_z$. That is to say, for instance, at $T=0$ a 3D chiral d-wave pairing with $\Delta_{\bk} =\Delta(k_x+ik_y)k_z$ shall exhibit the same edge current as a chiral p-wave with $\Delta_{\bk} =\Delta( k_x+ik_y)$, as is verified in Fig.~\ref{fig:3DFS} (barring some finite size effects in the calculation). This could be understood by noting that, at each $k_z$-plane, the former semiclassically behaves as a chiral p-wave with gap amplitude $\Delta k_z$, and that, in the BCS limit, the edge current is independent of the sign and amplitude of the pairing. Meanwhile, any 3D chiral superfluid with $L_z>\hbar$ shall exhibit vanishing OAM, just like their 2D counterparts. These conclusions also apply to models with anisotropic band dispersion, such as in a lattice model with anisotropic hopping parameters in Fig.~\ref{fig:3DFSani}. This holds special significance, given the recent speculation of 3D $(k_x+ik_y)k_z$-like chiral d-wave pairing in Sr$_2$RuO$_4$ \cite{Iida:19,LiYS:19}. If we take the absence of edge current as a given \cite{Kirtley:07,Hicks:10,Curran:14}, such a 3D chiral d-wave state is an equally unlikely candidate as the traditionally conceived 2D chiral p-wave pairing. The same goes for anisotropic 3D chiral p-wave models with horizontal line nodes~\cite{Roising:18}. As a side remark, in 3D models exhibiting nodal quasiparticle excitations, the current and the OAM could be more strongly suppressed at finite temperatures.

\section{2D chiral superfluids in BEC limit}
\label{sec:2Dbec}
Since Cooper pairs in BEC superfluids are usually conceived as tightly-bound molecules (Fig.~\ref{fig:illustration}), it is natural to assume a uniform OAM density distribution in the bulk of a BEC chiral superfluid. The absence of chiral edge modes as well as the absence of unpaired particles in a finite geometry~\cite{Tada:15} would seem to corroborate this assumption. Thus a total OAM of $\nu N\hbar/2$ is often expected for such a superfluid confined in a finite 2D disk, as was indeed confirmed in Ref.~\onlinecite{Tada:15}. On the other hand, in the single-particle perspective that we have taken for this work, the OAM is exclusively generated by spontaneous current -- which would have vanished in the absence of translation symmetry breaking. We are thus left to conjecture that the OAM in such a superfluid must again arise from boundary effects. This was indeed found to be true in an earlier study of the BEC chiral p-wave phase~\cite{Mizushima:08}. Therefore, the pictorial description in Fig.~\ref{fig:illustration} is somewhat misleading, and the OAM density shall vanish in the bulk. Here, we extend the study to higher-order chiral states. 

The BEC phase is acquired by setting $E_f<0$ in Eq.~(\ref{eq:Hamiltonian}). We follow the numerical BdG calculations employed in Ref.~\onlinecite{Tada:15} for BEC chiral superfluids on a 2D circular disk. The circular symmetry allows for the use of a convenient free particle basis with angular momentum quantum numbers, $l$. Some modest modifications are made here, in order to obtain a converging spatial profile of the physical quantities such as particle density and spontaneous current (see also App.~\ref{app3}). Specifically, instead of using the unregulated gap function $\Delta_{\bs k} =\Delta (k_x + ik_y)^\nu$ in Ref.~\onlinecite{Tada:15} which diverges at ultraviolet, we divide this pair potential by a term effectively of order $k^\nu$ to model the regulated gap function in Eq.~\ref{eq:Hamiltonian}. 

\begin{figure}
\subfigure{ \includegraphics[width=4cm]{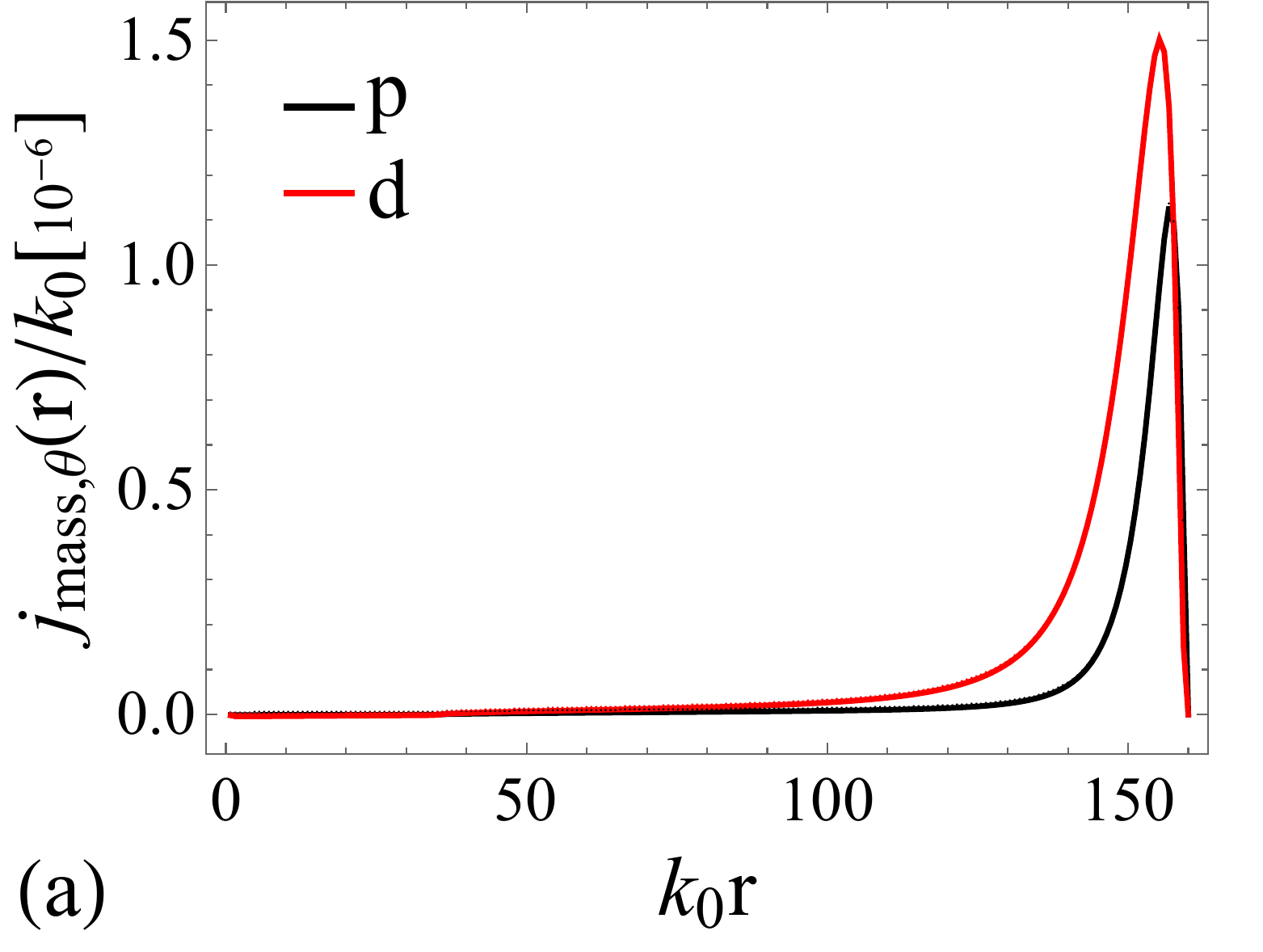}}
\subfigure{ \includegraphics[width=4cm]{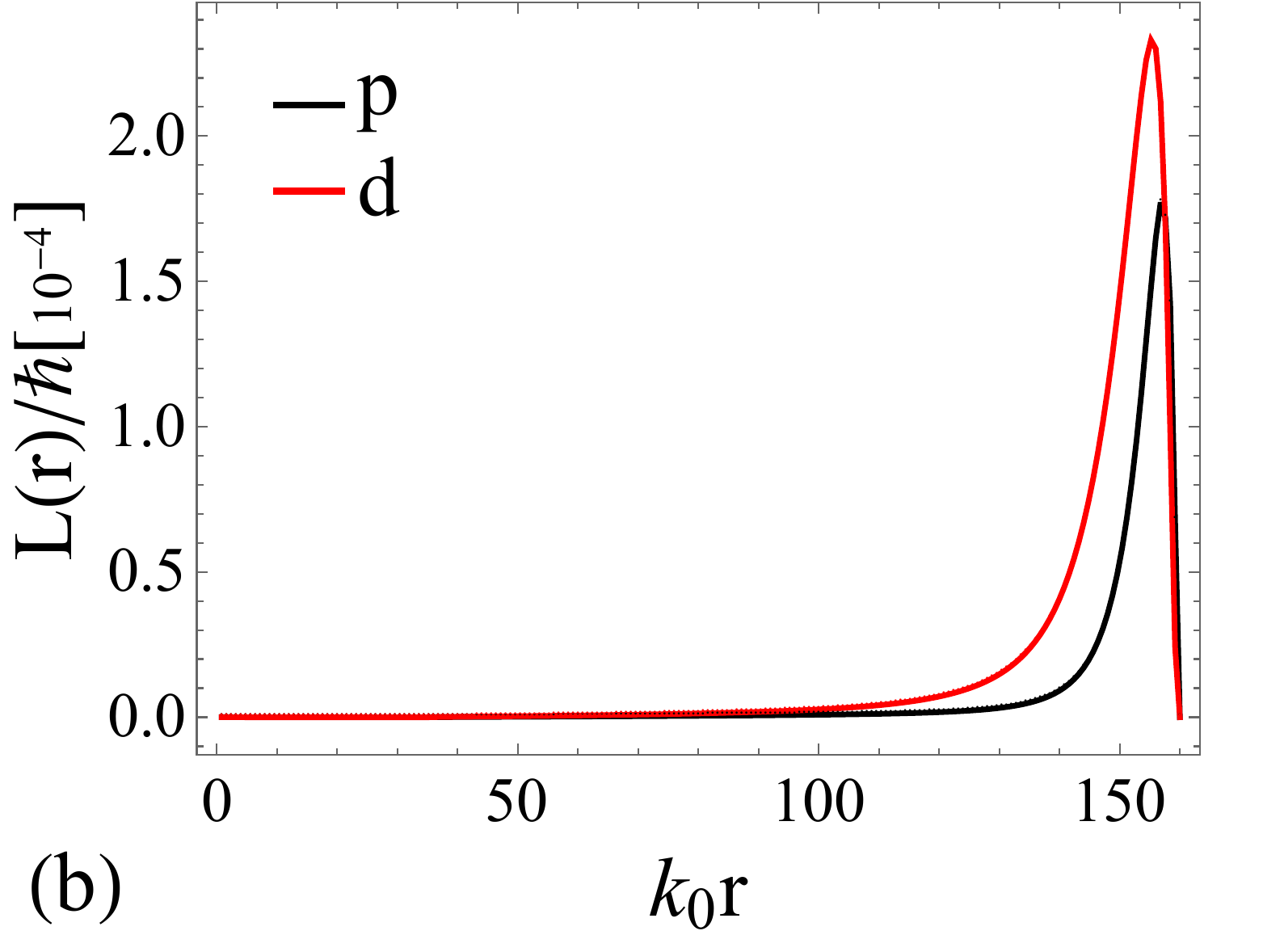}}
\subfigure{ \includegraphics[width=4cm]{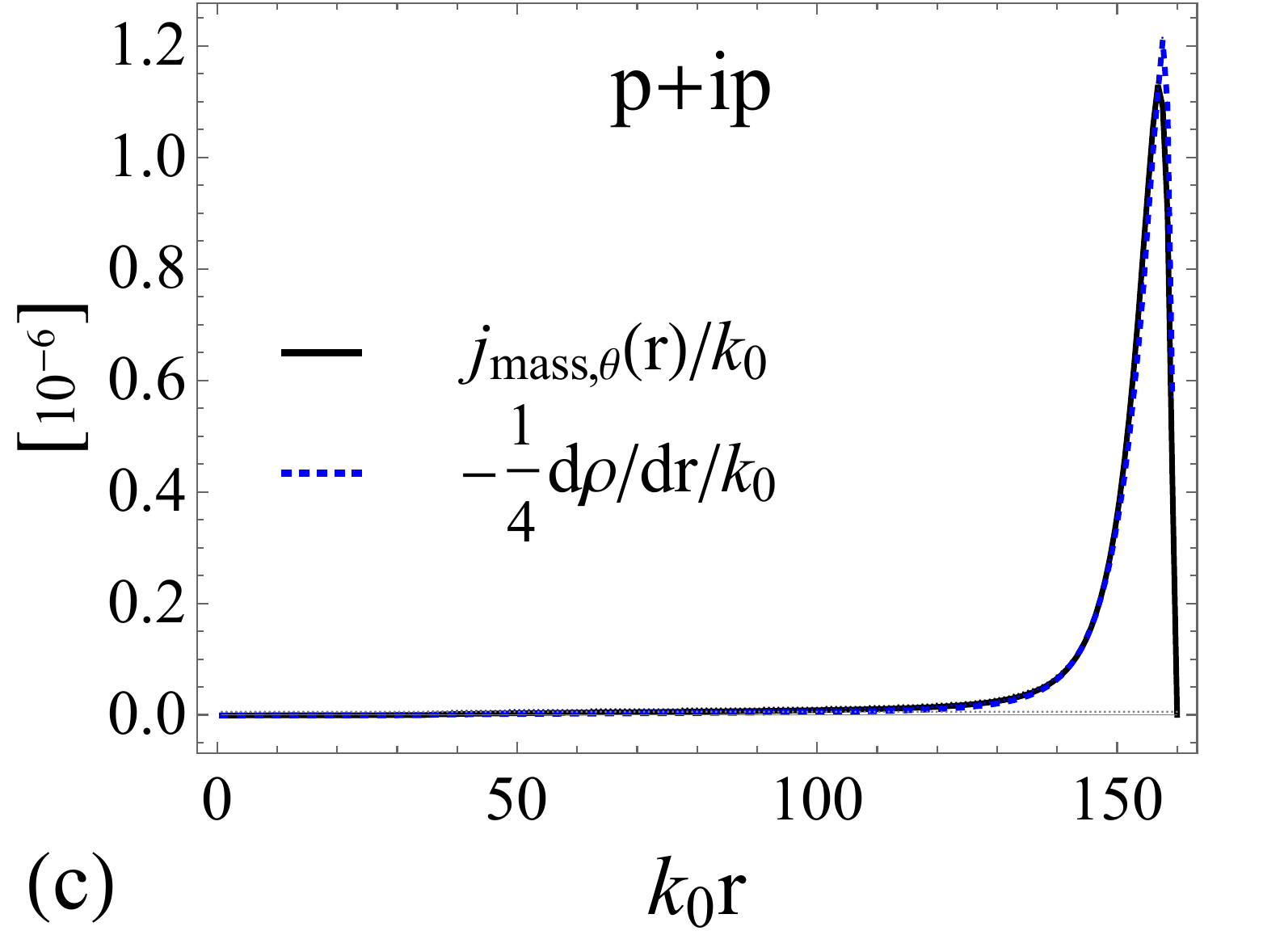}}
\subfigure{ \includegraphics[width=4cm]{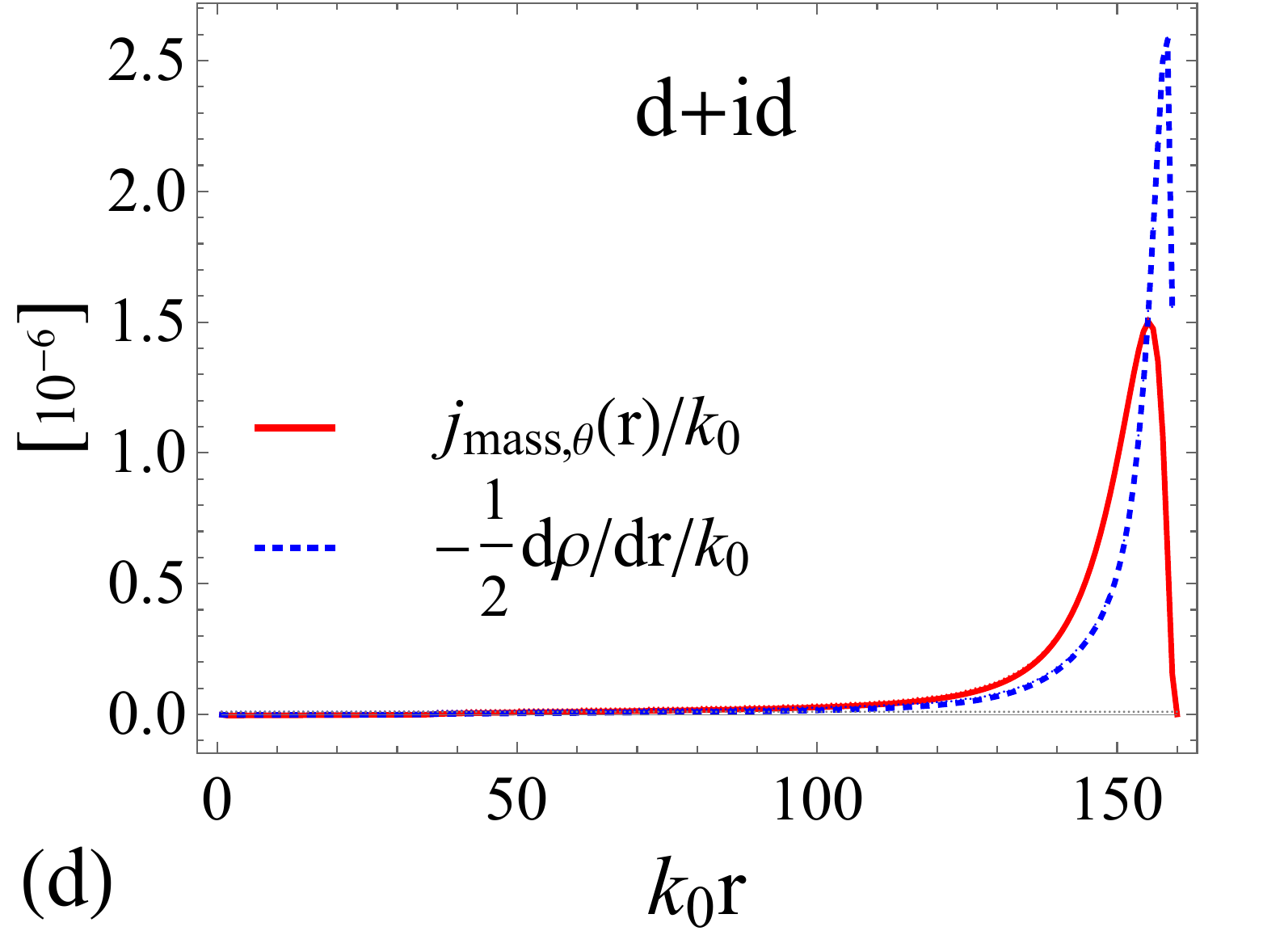}}
\caption{Spontaneous current and OAM in the BEC limit. (a) The distribution of azimuthal mass current, and (b) OAM density for chiral p-wave and d-wave superfluids in a $2$D disk. The mass current and density gradient for the two states are plotted in (c) and (d) for comparison, with $\mu=-0.025 E_0$, $k_0\Delta=0.005 E_0$ for p-wave and $k_0^2\Delta=0.005 E_0$ for d-wave superfluids, where $E_0=k_0^2/(2m)$, is taken as a fundamental unit of energy.}
\label{fig:bec}
\end{figure}

\begin{figure}
\includegraphics[width=6cm]{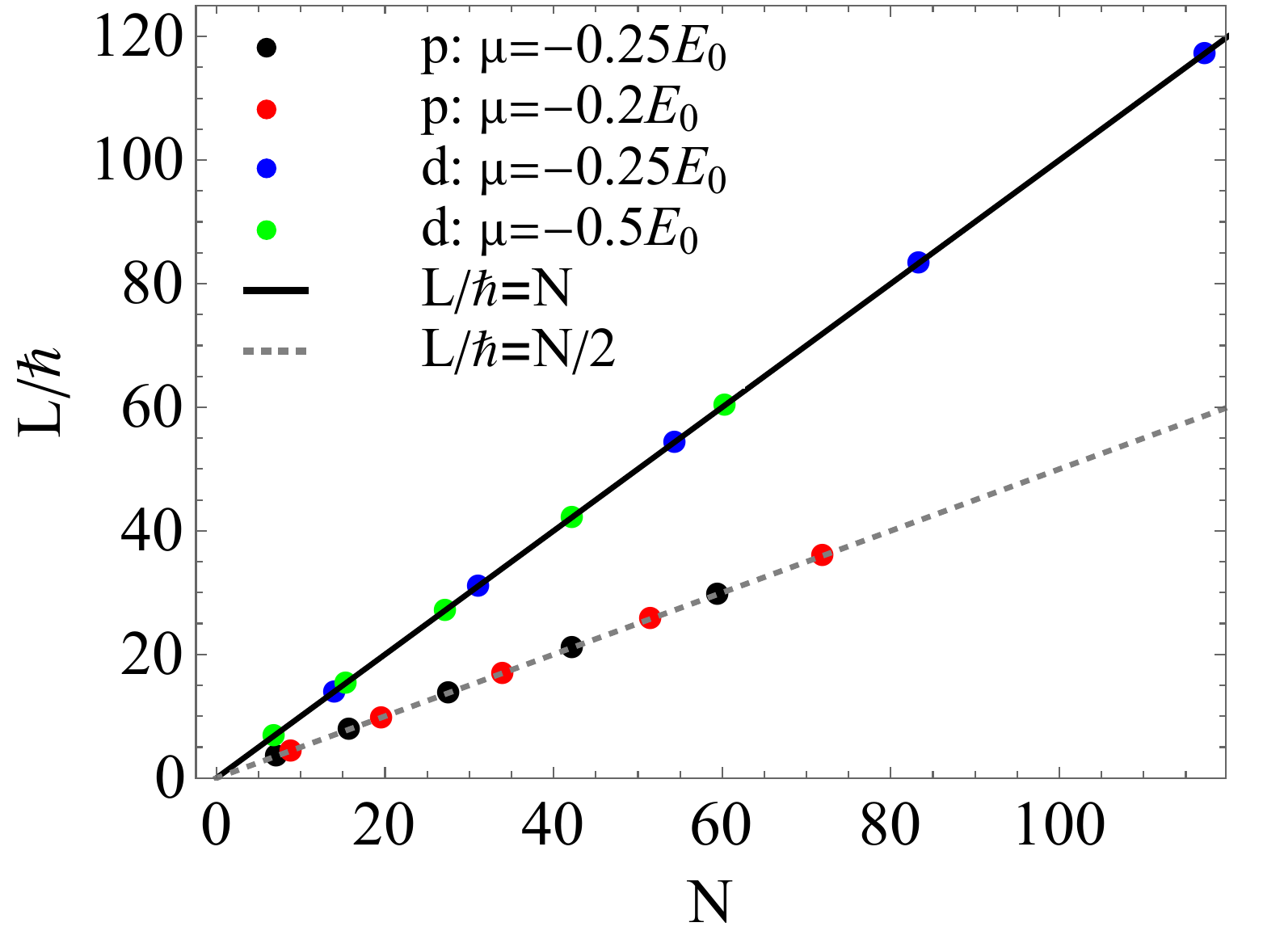}
\caption{Total OAM of BEC chiral p- and d-wave superfluids on a circular disk as a function of the particle number $N$. Variation of $N$ is achieved by varying $\Delta$ or $\mu$. For example, the black data points are obtained by varying $\Delta$ whiling keeping $\mu=-0.25E_0$.}
\label{fig:becLN}
\end{figure}

Main results of some representative calculations for chiral p- and d-wave models are presented in Figs.~\ref{fig:bec} and \ref{fig:becLN}. Despite the absence of chiral edge modes, thanks to the chirality of the Cooper pairing the scattering of quasiparticles at the boundary induces an overall spontaneous current. One most important observation in Fig.~\ref{fig:bec} is that spontaneous current emerges only at the boundary. The OAM distribution, given by $L_z(r) = 2\pi mr^2 j(r)$, is necessarily also confined to the boundary.

Note that, within the above formulation, the OAM each Cooper pair generates with respect to the disk center equals the relative OAM this pair carries. Since the relation $L_z^\text{tot} = \int_0^R L_z(r)dr=\nu N\hbar/2$ is exactly satisfied (see Fig.~\ref{fig:becLN}), and since all fermions are paired in BEC~\cite{Tada:15}, it is reasonable to conclude that the macroscopic OAM originates solely from the Cooper pair relative OAM. This contrasts with the BCS limit where some unpaired fermions exist at the edge and contribute a net OAM pointing in the opposite direction~\cite{Tada:15}.

Importantly, unlike in the BCS limit where the particle density remains roughly constant at the boundary while the spontaneous current decays over several coherence lengths, the BEC limit sees the particle density and the edge current varying over comparable length scales. More interestingly, in contrast to that in non-p-wave pairings, the current in chiral p-wave appears to follow the gradient of the particle density $\partial_r \rho(r)/4$ as indicated in Fig.~\ref{fig:bec} (c) and (d). Given that the Chern-Simon-like action is not operative in the BEC limit, the striking similarity to Eq.~(\ref{eq:CScurrent}) is puzzling and it lacks a formal explanation. As a final remark, our mean-field BdG calculations cannot account for the bosonic collective excitations which could have become the dominant source of fluctuations in the BEC superfluid -- effectively a boson system. Such fluctuations, unable to be captured in our BdG, induce a phase-coherence length scale distinct from the Cooper pair size \cite{Pistolesi:95,Andrenacci:03,Taylor:06}. While it goes beyond the scope of the present study, the influence of these fluctuations to the spontaneous current and the OAM is a problem worthy of further investigation.

\section{Summary}
\label{sec:summary}
Based on the semiclassical BdG theory, we studied the spontaneous edge current and OAM of chiral superfluids in finite geometry. Within this theory, the spontaneous current and OAM are both single-particle quantities closely tied to the quasiparticle scattering at the system boundary. Following several previous studies, we substantiated the semiclassical analysis which provides an intuitive explanation for the vanishing of OAM in 2D non-p-wave BCS chiral superfluids. The same analysis also describes well the physics in anisotropic chiral superconductors as well as 3D chiral superfluids and superconductors. Going to the BEC limit, the current and OAM density are also found to be confined to the boundary, in contrast to the na{\"i}ve expectation for uniformly distributed tightly-bound Cooper pair molecules. In brief, our study brings new understanding of the enigmatic {\it Angular Momentum Paradox} in chiral superfluids.

\section{Acknowledgements}
We would like to acknowledge extensive earlier collaborations on this subject with Masaki Oshikawa and Yasuhiro Tada (W.N.), and Catherine Kallin, Edward Taylor and Samuel Lederer (W.H.). These collaborations laid the foundation for the present work. We are also grateful to Shizhong Zhang and Zhongbo Yan for many valuable discussions. This work is supported in part by the NSFC under Grant No. 11704267 (W.N.), a start-up funding from Sichuan University under Grant No. 2018SCU12063 (W.N.), the NSFC under grant No. 11825404 (H.Y.) and No. 11904155 (W.H.), the National Key Research and Development Program of China under grant No. 2016YFA0301001 and No.2018YFA0305604 (H.Y.), the Strategic Priority Research Program of Chinese Academy of Sciences under Grant No.~XDB28000000, Beijing Municipal Science and Technology Commission under Grant No. Z181100004218001, Beijing Natural Science Foundation under Grant No. Z180010 (H.Y.), the Guangdong Provincial Key Laboratory under Grant No. 2019B121203002 and the C.N. Yang Junior Fellowship at Tsinghua University (W.H.).

\appendix 
\section{Edge state solution in BCS limit}
\label{app1}
In this appendix we illustrate the derivation of the edge states in 2D chiral p- and d-wave states. We begin with the following BdG equation, 
\begin{equation}
\begin{bmatrix}
-\frac{\bs \partial^2}{2m}-E_f   & \Delta \left( \frac{i\partial_x + \partial_y}{k_f} \right)^\nu \\
\Delta \left( \frac{i\partial_x - \partial_y}{k_f}\right)^\nu & \frac{\bs \partial^2}{2m}+E_f  
\end{bmatrix}  
\begin{bmatrix}
u_0(\bs r)\\
v_0(\bs r) 
\end{bmatrix}
 = \epsilon 
\begin{bmatrix}
u_0(\bs r)\\
v_0(\bs r) 
\end{bmatrix}
\label{eq:H0}
\end{equation}
Note that we have removed the $k$-dependence in the denominator of the gap function by replacing $k$ with $k_f$. As we note in the main text, this modification simplifies our analysis while keeping the essential property of the edge states unaltered. The short-wavelength component in the wavefunction can be integrated out by taking $[u_0(\bs r),v_0(\bs r)]^T =e^{i \bs k_f \cdot \bs r}[u^\prime (\bs r),v^\prime (\bs r)]^T$. Keeping only the leading order terms in each matrix element on the LHS of Eq. (\ref{eq:H0}), we arrive at the Andreev equation,
\begin{equation}
\begin{bmatrix}
-i\bs v_f\bs\cdot \bs \partial  & \Delta \left( \frac{k_{fx} + ik_{fy}}{k_f} \right)^\nu \\
\Delta \left( \frac{k_{fx} - ik_{fy}}{k_f} \right)^\nu & i\bs v_f\bs\cdot \bs \partial  
\end{bmatrix}  
\begin{bmatrix}
u^\prime (\bs r)\\
v^\prime (\bs r) 
\end{bmatrix}
 = \epsilon 
\begin{bmatrix}
u^\prime (\bs r)\\
v^\prime (\bs r) 
\end{bmatrix}
\label{eq:H1}
\end{equation}
where $v_f =k_f/m$ and $\bs k_f=(k_{fx},k_{fy})$. Note that the terms carrying $\bs \partial^2$ in the diagonal elements and those carrying $\partial_{x/y}$ in the off-diagonal elements are dropped to a good approximation, because both $u^\prime(\bs r)$ and $v^\prime(\bs r)$ contain only long-wavelength components which vary at length scales much longer than $k_f^{-1}$. We stress that this approximation is valid as long as the superconducting coherence length is much larger than $k_f^{-1}$, i.e. when $\Delta/E_f \ll 1$.  

In a half-infinite geometry with an ideal sharp edge parallel to $y$-axis, the translation symmetry along $y$ allows us to write $[u^\prime(\bs r),v^\prime(\bs r)]^T =\phi(x)^T = [u (x),v (x)]^T$, therefore,
\begin{equation}
(\hat{H}_\perp+ \hat{H}_\parallel ) \phi(x) = \epsilon \phi(x) \,,
\end{equation}
where, for chiral p-wave, 
\begin{eqnarray}
\hat{H}_{\perp} &=& -iv_{fx}\partial_x \sigma_3 + \Delta k_{fx}/k_f \sigma_1 \,, \nonumber \\
\hat{H}_{\parallel} &=& - \Delta k_{fy}/k_f \sigma_2 \,.
\label{eq:PwaveH}
\end{eqnarray}
and for chiral d-wave,
\begin{eqnarray}
\hat{H}_{\perp} &=& -iv_{fx}\partial_x \sigma_3 - 2\Delta k_{fx}k_{fy}/k_f^2 \sigma_2 \,, \nonumber \\
\hat{H}_{\parallel} &=& - \Delta (k^2_{fx}-k^2_{fy})/k^2_f \sigma_1 \,.
\label{eq:DwaveH}
\end{eqnarray}
At each $k_y \equiv k_{fy}$, $\hat{H}_\perp$ in Eq.~(\ref{eq:PwaveH}) [and separately in Eq.~(\ref{eq:DwaveH})] constitutes an effective 1D Dirac domain wall problem~\cite{Stone:04}, in which scenario the opposite momenta $k_{fx}$ and $-k_{fx}$ of the respective incident and reflected waves lead to opposite masses on the two sides of the fictitious domain wall, such as $\Delta k_{fx}/k_f$ and $-\Delta k_{fx}/k_f$ in the p-wave model. This domain wall binds a zero-energy mode according to the Jackiw-Rebbi theory~\cite{Jackiw:76,Goldstone:81}. Further, since in each case $\hat{H}_\perp$ exhibits a chiral symmetry and since $\hat{H}_\parallel$ happens to be proportional to the corresponding chiral operator, the bound state solution of $\hat{H}_\perp$ must correspond to an eigenstate of $\hat{H}_\parallel$. For chiral p-wave, the eigenvector is $(1,-i)^T/\sqrt{2}$ with energy dispersion $E_{k_y} = \Delta k_{fy}/k_f$, and the approximate full solution for Eq.~(\ref{eq:H0}) at $\nu=1$ can be shown to take the following form,
\begin{equation}
\phi_{k_y}(x,y) \propto \begin{pmatrix}
1 \\
-i
\end{pmatrix} \sin(k_{fx}x)e^{-\frac{\Delta}{v_f} x}e^{ik_{fy}y}  \,.
\end{equation}
Note that the choice of a particular eigenstate of $\hat{H}_\parallel$ has to comply with the boundary condition, or alternatively, with the Chern number of the chiral pairing as manifest in the number and chirality of chiral branches. For chiral d-wave, the eigenvector is $(1,\pm 1)^T/\sqrt{2}$ with energy dispersion $E_{k_y} =\pm \Delta (k^2_{fx}-k^2_{fy})/k^2_f$ for the two chiral branches in Fig.~\ref{fig:edgeDisp} b. For completeness, we write down directly the chiral edge dispersion in the chiral f-wave state as follows (Fig. \ref{fig:edgeDisp} c),
\begin{eqnarray}
E_{k_y} &=& \left\{ \begin{array}{cc} 
                -\Delta \frac{3k_f^2k_y-4k_y^3}{k_f^3}\,, & ~k_y\in(-k_f,-\frac{k_f}{2}] \\
                \Delta\frac{3k_f^2k_y-4k_y^3}{k_f^3}\,, & ~k_y\in(-\frac{k_f}{2},\frac{k_f}{2}] \\
                -\Delta\frac{3k_f^2k_y-4k_y^3}{k_f^3}\,, & ~k_y\in(\frac{k_f}{2},k_f] \,. \\
                \end{array} \right. \nonumber \\
\end{eqnarray}

We see from above that the bound states are charge-neutral with equal-amplitude particle and hole composition. However, since the bound state solutions were obtained under the approximation $\Delta/E_f \ra 0$, it is not obvious that the charge-neutrality is protected by symmetry. To this end, we turn back to Eq.~(\ref{eq:H0}) to look for exact properties. It is instructive to perform a partial Fourier transformation along the $y$-direction, which yields the following matrix,
\begin{equation}
\begin{bmatrix}
-\frac{\partial_x^2}{2m}-E_f-\frac{k_y^2}{2m}  & \Delta \left( \frac{-i\partial_x + ik_y}{k_f} \right)^\nu \\
\Delta \left( \frac{-i\partial_x - ik_y}{k_f} \right)^\nu & \frac{\partial_x^2}{2m}+E_f+\frac{k_y^2}{2m}
\end{bmatrix}  \,.
\end{equation}
In the case of $\nu=1$, the matrix can be similarly decomposed into $\hat{H}_\perp$ and $\hat{H}_\parallel$, which anticommute with each other and where $\hat{H}_\parallel =-\Delta \frac{k_y}{k_f}\sigma_2$ has no $x$-dependence. Hence the solution of the chiral edge modes in a chiral p-wave superfluid is given by an eigenvector of $\sigma_2$, and the charge neutrality is thus protected by an exact chiral symmetry. By contrast, for $\nu>1$, the corresponding $\hat{H}_\perp$ generically does not exhibit any chiral symmetry and it does not anticommute with the $x$-independent $\hat{H}_\parallel$, thus the charge-neutrality of the bound states we obtained above is not exact. The correction turns out to be order $\Delta/E_f$ as one may infer from our preceding approximation. The same conclusion applies to all higher chirality pairings. Following the same analyses, in the lattice chiral p-wave models where each of the individual component has both $k_x$ and $k_y$ dependence, such as $\Delta_{\bs k} \sim \sin k_x \cos k_y + i \cos k_x \sin k_y$, the edge modes similarly receive some order $\Delta/E_f$ corrections, as we verify in the next section. 

\begin{figure}
\subfigure{\includegraphics[width=7cm]{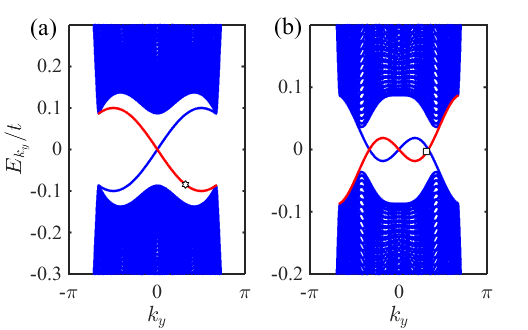} }
\subfigure{\includegraphics[width=7cm]{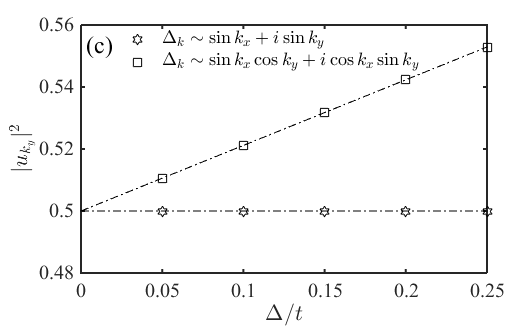} }
\subfigure{\includegraphics[width=7cm]{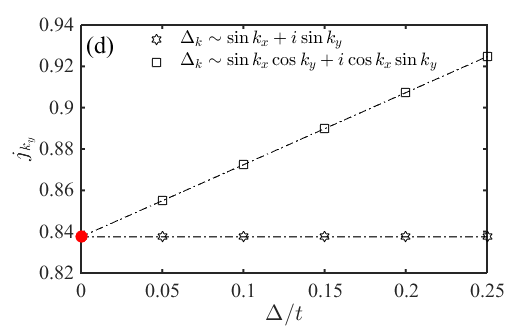} } 
\caption{(color online) Results of the BdG calculations on a stripe geometry for a square lattice chiral p-wave models with only nearest-neighbor hopping $t$. Shown in (a) and (b) are the low energy spectra at $\Delta=0.1t$, where the states marked in red represent the chiral modes localized at one of the edges. The pairing function acquires the form $\Delta_{\bf k} = \Delta(\sin k_x - i\sin k_y)$ in (a) and $\Delta_{\bf k} = \Delta(\sin k_x\cos k_y +i\cos k_x \sin k_y)$ in (b). The chemical potential is set at $\mu=-t$. The open hexagon in (a) and open square in (b) represent mark edge states at the same wavevector $k_y=0.316\pi$. (c) The $\Delta$-dependence of weight of one of the Nambu spinor components of the edge state with wavevector $k_y=0.316\pi$. (d) The $\Delta$-dependence of the total current carried by the edge state with $k_y=0.316\pi$. The red dot denotes the theoretical prediction in the limit $\Delta/t \ra 0$ based on the lattice version of Eq.~(\ref{eq:Jky}).}
\label{fig:chiralPbdg}
\end{figure}

\begin{figure}
\subfigure{\includegraphics[width=6.8cm]{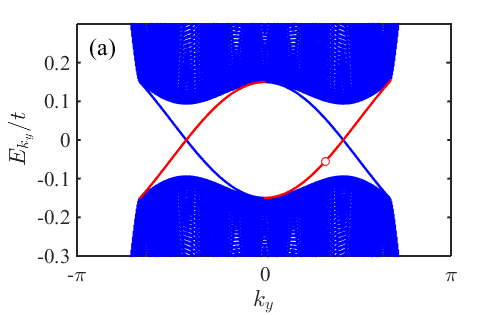} }
\subfigure{\includegraphics[width=6.8cm]{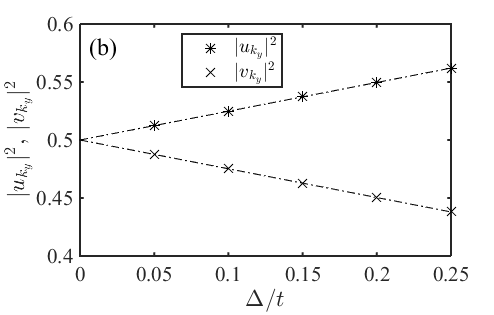} }
\subfigure{\includegraphics[width=6.8cm]{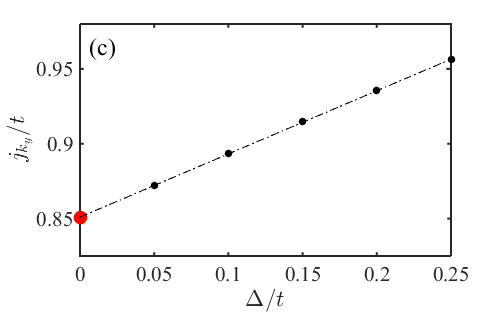} } 
\caption{(color online) Results of the BdG calculations on a stripe geometry for a square lattice chiral d-wave pairing with only nearest-neighbor hopping $t$ and $\Delta_{\bs k} = \Delta(\cos k_x - \cos k_y + 2i\sin k_x \sin k_y)$. The chemical potential is set at $\mu=-t$. (a) Low energy spectra (with $\Delta=0.1t$) where the states marked in red represent the chiral edge modes at one of the edges. The open circle highlights an edge state with wavevector $k_y=0.324\pi$. (b) The $\Delta$-dependence of weight of the particle and hole components of the edge state with $k_y=0.324\pi$. (c) The $\Delta$-dependence of the total current carried by the edge state with $k_y=0.324\pi$. The red dot denotes the theoretical prediction in the limit $\Delta/t \ra 0$ based on the lattice version of Eq.~(\ref{eq:Jky}).}
\label{fig:chiralDbdg}
\end{figure}

\section{Lattice BdG calculations}
\label{app:nuBdG}
In this section we present our numerical BdG calculations on 2D square lattice models. Calculations on the 3D cubic lattice models can be generalized straightforwardly. The BdG Hamiltonian is a sum of kenetic and pairing terms, 
\begin{equation}
H= H_t + H_\Delta \,.
\end{equation}
The kinetic term is given by, 
\begin{equation}
H_t= -\sum_{m,n,\sigma} t_{mn} c^\dagger_{m,\sigma}c_{n,\sigma} -\mu \sum_{m,\sigma} c^\dagger_{m,\sigma}c_{m,\sigma} \,
\end{equation}
where $\sigma$ denote the spin species and $t_{mn}$ the hopping between site $m$ and site $n$. If only nearest neighbor hopping $t$ is considered, this term in momentum space follows as $\xi_k = -2t (\cos k_x + \cos k_y) -\mu$. The pairing term must acquire the $k_x+ik_y$ symmetry in chiral p-wave and $k_x^2-k_y^2+2ik_xk_y$ in the chiral d-wave model, etc. For example, in the simplest chiral p-wave model with $\Delta_k = \Delta_0 (\sin k_x + i \sin k_y)$, it is realized by,
\begin{eqnarray}
H_\Delta&=& i \Delta_0 \sum_m ( c_{m,\ua}c_{m+\hat{x},\da} -c_{m+\hat{x},\ua}c_{m,\da} ) + H.c. \nonumber \\
 &+& \Delta_0 \sum_m ( c_{m,\ua}c_{m+\hat{y},\da} -c_{m+\hat{y},\ua}c_{m,\da} ) +H.c. \,,
\end{eqnarray}
where $\hat{x}$ and $\hat{y}$ are vectors of unit length in the x- and y-directions. In this expression the phase factor $i$ is responsible for the $\pi/2$ phase difference between the x- and y-components of the p-wave pairing. 


The particle current of the spin-$\sigma$ fermion flowing from site $n$ to site $m$ is given by \cite{Huang:15},  
\begin{equation}
J_{mn,\sigma} = id_{mn}t_{mn}(c^\dagger_{m,\sigma} c_{n,\sigma} -  c^\dagger_{n,\sigma} c_{m,\sigma})
\end{equation}
where $d_{mn}$ is the length of the bond connecting $m$ and $n$. Notice this current operator can be obtained from a standard Peierls substitution. 

The actual calculation is performed on a cylindrical geometry with open boundaries in the x-direction and a periodic boundary condition in y-direction. Hence momentum $k_y$ is a good quantum number. We hence perform a Fourier transformation in the y-direction and keep the real space site indices along x. The BdG Hamiltonian and the current operator can then be written for each value of $k_y$. We numerically diagonalize the Hamiltonian at each $k_y$ and compute the expectation value of the current. The current carried by any individual Bogoliubov quasiparticle mode can also be evaluated. 

Figures \ref{fig:chiralPbdg} and \ref{fig:chiralDbdg} show the representative results of our numerical calculations for chiral p-wave and chiral d-wave models, respectively. In the simple chiral p-wave $\Delta_{\bs k} \sim \sin k_x - i\sin k_y$, we see that the edge states are exactly charge neutral with particle (or hole) amplitude of $|u_{k_y}|=0.5$ [Fig.~\ref{fig:chiralPbdg} (c)]. For the anisotropic p-wave pairing $\Delta_{\bs k} \sim \sin k_x\cos k_y +i\cos k_x \sin k_y$ where each component has both $k_x$ and $k_y$ dependence, the particle (or hole) amplitude and the current carried by the edge mode exhibit linear-$\Delta$ corrections [Fig.~\ref{fig:chiralPbdg} (c) and (d)]. Nevertheless,  in the limit $\Delta/t\ra 0$, the edge states approach charge neutrality and the current approaches the predicted value given by the lattice version of Eq.~(\ref{eq:Jky}): $j_{k_y} = \partial_{k_y}\xi_{\bs k}/2$ where $\xi_{\bs k} = -2t(\cos k_x + \cos k_y)-\mu$ in the present calculation, independent of the detailed structure of the p-wave pairing function or the edge dispersion. Notice that the highlighted edge states in Fig.~\ref{fig:chiralPbdg} (a) and (b) have the same wavevector $k_y$ but are characterized by distinct group velocities. Further, we checked that at finite $\Delta$ the correction in the current carried by an individual mode in anisotropic p-wave originates purely from the correction in the particle (or hole) amplitude of the wavefunction, and that it has nothing to do with the group velocity of the edge mode. The edge states in the chiral d-wave model exhibit similar behavior as those in the above anisotropic p-wave model, as can be seen in Fig.~\ref{fig:chiralDbdg}.

\section{BdG calculations in the BEC limit}
\label{app3}

We consider the two-dimensional chiral superfluids confined in a circular well with a specular wall, in the framework of BdG Hamiltonian. The $\ve{d}$-vector has no variation near the boundary and takes the form $\ve{d}={(0,0,d_z)}$ everywhere. We consider the mean-field Hamiltonian 
$\hat{H}=\int d^2x\psi^{\dagger}_{\sigma}[(p_x^2+p_y^2)/2m+V-\mu]
\psi_{\sigma}
+\int d^2x\psi_{\uparrow}^{\dagger}\Delta (p_x+ip_y)^{\nu}/|p|^{\nu}
\psi_{\downarrow}^{\dagger}+({\rm H.c.}),$
where $p_{j}=-i\partial/\partial x_j$,
$m$ is the fermion mass, and $\mu$ is the chemical potential.
The confining potential $V(r)$ is chosen to be
$V(r<R)=0$ and $V(r>R)=\infty$ with a radius $R$ for infinite circular well. 

The circular geometry allows for an expansion of the field operators in the angular momentum basis
$\psi_{\sigma}(\ve{r})=\sum_{nl}c_{nl\sigma}\varphi_{nl}(\ve{r})$
where $\varphi(\ve{r})$ is a solution of the equation
$[(p_x^2+p_y^2)/2m+V(r)-\mu]\varphi_{nl}(\ve{r})=\varepsilon_{nl}
\varphi_{nl}(\ve{r})$.
Then the Hamiltonian becomes
\begin{align}
\hat{H}&=\sum_{l}\sum_{nn^{\prime}}
\left[
\begin{array}{c}
c_{n,l+\nu,\uparrow}^{\dagger}\\ c_{n,-l,\downarrow}
\end{array}\right]^T\notag \\
&\quad \times \left[
\begin{array}{cc}
\varepsilon_{n,l+\nu}\delta_{nn^{\prime}} & \Delta_{nn^{\prime}}^{(l)}\\
\Delta_{n^{\prime}n}^{(l)\ast}& -\varepsilon_{n,-l}\delta_{nn^{\prime}}
\end{array}\right]
\left[
\begin{array}{c}
c_{n^{\prime},l+\nu,\uparrow}\\ c_{n^{\prime},-l,\downarrow}^{\dagger}
\end{array}\right],
\label{eq:Ham}
\end{align}
where $\Delta_{nn^{\prime}}^{(l)}=\int \varphi_{n,l+\nu}^{\ast}
\Delta  (p_x+ip_y)^{\nu}/|p|^{\nu} \varphi_{n',-l}^{\ast}$,
where $|p|^{\nu}$ is introduced for converging spatial profile within a reasonable cutoff.

We evaluate the ground state expectations of the physical quantities. The particle density is given by $\rho(\ve{r}) =\langle \psi^{\dag}(\ve{r}) \psi (\ve{r}) \rangle$, and the particle current by $\ve{j} =\langle [ \psi^{\dag} (-i \ve{\nabla} \psi) + (i \ve{\nabla} \psi^{\dag})\psi] \rangle/2m$. Note that the continuity equation $\partial \rho/\partial t+ \ve{\nabla} \cdot \ve{j} =0$ is satisfied. Since no current flows in the radial direction in a disk geometry, the current can be re-expressed as $j_{\theta}(r)=\langle [\psi^{\dag} (-i \hbar \frac{\partial}{\partial \theta} \psi)+ (i \hbar \frac{\partial}{\partial \theta}\psi^{\dagger}) \psi]\rangle/(2mr)$. The distribution of the orbital angular momentum is given by $L(\ve{r}) = \langle \psi^{\dag}(\ve{r}) (-i \hbar \frac{\partial}{\partial \theta})\psi (\ve{r}) \rangle$, which implies $L(r)=m r j_{\theta}(r)$. In other words, the OAM originates entirely from the spontaneous current.


\begin{thebibliography}{99}
\bibitem{Anderson:61} P.W. Anderson and P. Moreal, Phys. Rev. {\bf 123}, 1911 (1961).
\bibitem{Leggett:75} A.J. Leggett, Rev. Mod. Phys. {\bf 47}, 331 (1975).
\bibitem{Anderson:73} P.W. Anderson and W.F. Brinkman, Phys. Rev. Lett. {\bf 30}, 1108 (1973). 
\bibitem{LeggettBook} A.J. Leggett, {\it Quantum Liquids: Bose condensation and Cooper pairing in condensed-matter systems}, Oxford Graduate Texts (2006). 
\bibitem{Mizushima:16} T. Mizushima, Y. Tsutsumi, T. Kawakami, M. Sato, M. Ichioka, and K. Machida, J. Phys. Soc. Jpn. {\bf 85}, 022001 (2016), and references therein. 
\bibitem{Volovik:75} G.E. Volovik, JETP Letters {\bf 22}, 108 (1975).
\bibitem{Mermin:75} N.D. Mermin and T-L. Ho, Phys. Rev. Lett. {\bf 36}, 594 (1975). 
\bibitem{Cross:77} M.C. Cross, J. Low Temp. Phys. {\bf 26}, 165 (1977). 
\bibitem{Ishikawa:77} M. Ishikawa, Prog. Theo. Phys. {\bf 57}, 1836 (1977). 
\bibitem{McClure:79} M.G. McClure and S. Takagi, Phys. Rev. Lett. {\bf 43}, 596 (1979).
\bibitem{Mermin:79} N.D. Mermin and P. Muzikar, Phys. Rev. B {\bf 21}, 980 (1980). 
\bibitem{Volovik:81} G.E. Volovik and V.P. Mineev, Sov. Phys. JETP {\bf 54}, 524 (1981). 
\bibitem{Kita:96} T.Kita. J. Phys. Soc. Jpn. {\bf 65}, 664 (1996).
\bibitem{Volovik:88} G.E. Volovik, Sov. Phys. JETP {\bf 67}, 1804 (1988). 
\bibitem{Matsumoto:99} M. Matsumoto and M. Sigrist, J. Phys. Soc. Jpn. {\bf 68}, 994 (1999).
\bibitem{Furusaki:01} A. Furusaki, M. Matsumoto, and M. Sigrist, Phys. Rev. B {\bf 64}, 054514 (2001).
\bibitem{Stone:04} M. Stone and R. Roy, Phys. Rev. B {\bf 69}, 184511 (2004).
\bibitem{Stone:08} M. Stone and I. Anduaga, Ann. Phys. (Amsterdam) {\bf 323}, 2 (2008). 
\bibitem{Sauls:11} J.A. Sauls, Phys. Rev. B {\bf 84}, 214509 (2011).
\bibitem{Huang:14} W. Huang, E. Taylor, and C. Kallin, Phys. Rev. B {\bf 90}, 224519 (2014).
\bibitem{Tada:15} Y. Tada, W. Nie, and M. Oshikawa, Phys. Rev. Lett. {\bf 114}, 195301 (2015).
\bibitem{Volovik:14} G.E. Volovik, JETP Lett. {\bf 100}, 742 (2014). 
\bibitem{Ojanen:16} T. Ojanen, Phys. Rev. B {\bf 93}, 174505 (2016).
\bibitem{Suzuki:16} S-I. Suzuki and Y. Asano, Phys. Rev. B {\bf 94}, 155302 (2016).
\bibitem{WangX:18} X. Wang, Z. Wang and C. Kallin, Phys. Rev. B {\bf 98}, 094501 (2018). 
\bibitem{Prem:17} A. Prem, S. Moroz, V. Gurarie, and L. Radzihovsky, Phys. Rev. Lett. {\bf 119}, 067003 (2017). 
\bibitem{Volovik:85} G.E. Volovik and L.P. Gor'kov, Sov. Phys. JETP {\bf 61}, 843 (1985).
\bibitem{Sigrist:89} M. Sigrist, T.M. Rice, and K. Ueda, Phys. Rev. Lett. {\bf 63}, 1727 (1989).
\bibitem{Kirtley:07} J.R. Kirtley, C. Kallin, C.W. Hicks, E.-A. Kim, Y. Liu, K.A. Moler, Y. Maeno, K.D. Nelson, Phys. Rev. B {\bf 76}, 014526 (2007).
\bibitem{Hicks:10} C.W. Hicks, J. R. Kirtley, T.M. Lippman, N. C. Koshnick, M. E. Huber, Y. Maeno, W.M. Yuhasz, M. B. Maple, and K. A. Moler, Phys. Rev. B {\bf 81}, 214501 (2010).
\bibitem{Curran:14} P. J. Curran, S. J. Bending, W. M. Desoky, A. S. Gibbs, S. L. Lee, and A. P. Mackenzie, Phys. Rev. B {\bf 89}, 144504 (2014).
\bibitem{Huang:15} W. Huang, S. Lederer, E. Taylor, C. Kallin, Phys. Rev. B {\bf 91}, 094507 (2015).
\bibitem{Tada:15b} Y. Tada, Phys. Rev. B {\bf 92}, 104502 (2015).
\bibitem{Ashby:09} P.E.C. Ashby and C. Kallin, Phys. Rev. B {\bf 79}, 224509 (2009).
\bibitem{Lederer:14} S. Lederer, W. Huang, E. Taylor, S. Raghu, and C. Kallin, Phys. Rev. B {\bf 90}, 134521 (2014).
\bibitem{Bouhon:14} A. Bouhon and M. Sigrist, Phys. Rev. B {\bf 90}, 220511(R) (2014).
\bibitem{Scaffidi:15} T. Scaffidi and S.H. Simon, Phys. Rev. Lett. {\bf 115}, 087003 (2015).
\bibitem{Tada:18} Y. Tada, Phys. Rev. B {\bf 97}, 214523 (2018).
\bibitem{footnote1} Although the semiclassical analysis assumed constant chiral order parameter components, the resultant quasiparticle states would nevertheless encode the peculiar behavior of the two components at the boundary. This can be seen by reconstructing the spatially resolved individual superconducting order parameter components using the obtained quasiparticle states, both in our numerics and in Ref.~\onlinecite{Sauls:11}.
\bibitem{Volovik:95} G.E. Volovik, JETP Lett. {\bf 61}, 958 (1995).
\bibitem{Braunecker:05} B. Braunecker, P. A. Lee, and Z. Wang, Phys. Rev. Lett. {\bf 95}, 017004 (2005).
\bibitem{Zhang:18} J-L. Zhang, W. Huang and D-X. Yao, Phys. Rev. B {\bf 98}, 014511 (2018). 
\bibitem{Etter:18} S.B. Etter, A. Bouhon and M. Sigrist, Phys. Rev. B {\bf 97}, 064510 (2018).
\bibitem{Goryo:99} J. Goryo and K. Ishikawa, Phys. Lett. A {\bf 260}, 294 (1999).
\bibitem{Roy:08} R. Roy and C. Kallin, Phys. Rev. B {\bf 77}, 174513 (2008).
\bibitem{Lutchyn:08} R.M. Lutchyn, P. Nagornykh, and V.M. Yakovenko, Phys. Rev. B {\bf 77}, 144516 (2008).
\bibitem{Hoyos:14} C. Hoyos, S. Moroz, and D.T. Son, Phys. Rev. B {\bf 89}, 174507 (2014).
\bibitem{Yoshiaka:18} N. Yoshioka, Y. Imai and M. Sigrist, J. Phys. Soc. Jpn. {\bf 87}, 124602 (2018).
\bibitem{Iida:19} K. Iida, M. Kofu, K. Suzuki, N. Murai, S. Ohira-Kawamura, R. Kajimoto, Y. Inamura, M. Ishikado, S. Hasegawa, T. Masuda, Y. Yoshida, K. Kakurai, K. Machida, S. Lee, J. Phys. Soc. Jpn {\bf 89}, 053702 (2020).
\bibitem{LiYS:19} Y.-S. Li, N. Kikugawa, D.A. Sokolov, F. Jerzembeck, A.S. Gibbs, Y. Maeno, C.W. Hicks, M. Nicklas, A.P. Mackenzie, arXiv:1906.07597. 
\bibitem{Roising:18} H.S. R\o{}ising, F. Flicker, T. Scaffidi, and S.H. Simon, Phys. Rev. B {\bf 98}, 224515 (2018).
\bibitem{Mizushima:08} T. Mizushima, M. Ichioka, and K. Machida, Phys. Rev. Lett. {\bf 101}, 150409 (2008). 
\bibitem{Pistolesi:95} F. Pistolesi and G.C. Strinati, Phys. Rev. B {\bf 53}, 15168 (1996). 
\bibitem{Andrenacci:03} N. Andrenacci, P. Pieri, and G.C. Strinati, Phys. Rev. B {\bf 68}, 144507 (2003). 
\bibitem{Taylor:06} E. Taylor, A. Griffin, N. Fukushima, and Y. Ohashi, Phys. Rev. A {\bf 74}, 063626 (2006).
\bibitem{Goldstone:81} J. Goldstone and F. Wilczek, Phys. Rev. Lett. {\bf 47}, 986 (1981).
\bibitem{Jackiw:76} R. Jackiw and C. Rebbi, Physical Review D {\bf 13}, 3398 (1976).


\end{thebibliography}
\end{document}